\documentclass[aps,pre,longbibliography,10pt,superscriptaddress,showpacs,showkeys,twocolumn]{revtex4-1}
\usepackage{multirow}
\usepackage{float}
\usepackage{graphicx}\usepackage{color,epsfig}
\usepackage{amssymb,amsmath}
\usepackage[ansinew]{inputenc}
\usepackage{bm}
\usepackage{enumerate}
\usepackage[T1]{fontenc}
\usepackage{array}
\usepackage{setspace}

\renewcommand{\vec}[1]{\boldsymbol{#1}}

\definecolor{darkgreen}{rgb}{0.0,0.6,0.0}

\begin{document}

\title{Energy localization and shape transformations in semiflexible polymer rings}

\author{Yu.B. Gaididei}
\affiliation{Bogolyubov Institute for Theoretical Physics, Metrologichna str. 14 B, 03143 Kiev, Ukraine}

\author{J.F.R. Archilla}
\affiliation{Grupo de F\'{i}sica No Lineal, Universidad de Sevilla, ETSI Inform\'{a}tica, A. Reina Mercedes s/n, 41012-Sevilla, Spain}

\author{V.J. S\'{a}nchez-Morcillo}
\affiliation{Instituto de Investigaci\'{o}n para la Gesti\'{o}n Integrada de las Zonas Costeras, Universidad Polit\'{e}cnica de Valencia, Paranimf 1, 46730 Grao de Gandia, Spain}

\author{C. Gorria}
\affiliation{Department of Applied Mathematics and Statistics, University of the Basque Country, E-48080 Bilbao, Spain}


\begin{abstract}
Shape transformations in driven and damped  molecular chains are considered.  Closed  chains of weakly coupled molecular subunits under the action of
spatially homogeneous time-periodic external  field are studied. The coupling between  the internal excitations  and the bending degrees of  freedom of the chain modifies the local bending rigidity
  of the chain. In the absence of driving the array takes a circular shape. When the energy pumped into the system exceeds some critical value the chain undergoes a non-equilibrium phase transition: the circular shape of the aggregate becomes unstable and the chain takes the shape of an ellipse or, in general, of a polygon. The excitation energy distribution becomes spatially nonuniform: it localizes in such places where the
chain is more flat. The weak interaction of the chain with a flat surface restricts the dynamics to a flat manifold.
\end{abstract}

\maketitle

\section{Introduction}
Nonlinear localization phenomena are widely recognized as key to
understanding the excitation dynamics in many physical and
technological aspects such as light propagation, charge and energy
transport in condensed-matter physics, or dynamics of micromechanical
oscillator arrays \cite{christ,christ2,souk,satoR}.  Recent advances in manufacturing of micro-
and nano-electromechanical systems \cite{craig} have made it
possible to fabricate externally actuated resonant nanostructures
and study intrinsic localized state formation in driven
micro-mechanical cantilever arrays \cite{christ,christ2,souk,satoR}.
Electric actuation is the most common actuation  method because of
its simplicity and high efficiency. However, other actuation
methods including thermal mechanical stresses,  magnetic fields
and  optical excitation  are also used \cite{blick,ilic}.

It is well known that in conservative
systems of nonlinear oscillators the modulational instability of
band edge plane waves leads to creation of spatially localized
states. However, as it was shown quite recently \cite{mani} in
damped and driven lattices intrinsic localized states appear via a
new instability mechanism, different from the modulational
instability. New aspects in nonlinear energy localization appear
in systems with complicated geometry. Nonlinear whispering gallery
modes for a nonlinear Maxwell equation in a microdisk were
investigated in \cite{harayama}, and the excitation of
whispering-gallery-type electromagnetic modes by a moving fluxon
in an annular Josephson junction was found in  \cite{ustinov}.
Localization of linear and nonlinear excitations in parabolically
curved waveguides was studied in \cite{panos}. A curved chain
of nonlinear oscillators was considered in \cite{curve} and
it was shown that the interplay of curvature and nonlinearity
leads to a symmetry breaking when an asymmetric stationary state
becomes energetically more favorable than a symmetric stationary
state. The interaction of classical anharmonic localized modes with
geometry was considered in Refs.~\cite{archilla2001a,archilla2001b,archilla2002,larsen2004a,larsen2004b}.
Another recent example of localization in curved geometries was reported in \cite{victor2014,victor2015}, where
spatial instabilities of a circular ring of coupled pendula
parametrically driven by a vertical harmonic force are discussed. Normal oscillation modes \cite{victor2015}
(breathing, dipole, quadrupole) and localized patterns of different types \cite{victor2014} (breathers
and kinks) are predicted and observed in such mechanical system. The analogy between the considered discrete
mechanical system and a gas bubble cavitating under the action of an acoustic field
was also established.

The bulk of theoretical results has been achieved for arrays of
nonlinear oscillators with fixed geometry: linear chains and
two-dimensional lattices. Until recently  there have been few
theoretical and numerical studies of nonlinear excitations in
systems with flexible geometry. Many types of biomolecules as polymers or DNA chains belong to this category.
Conformational dynamics with account of coupling between
the internal and mechanical degrees of freedom was studied in
\cite{ming}. It was found that the presence of nonlinear
excitations may cause the buckling and collapse instabilities of
an initially straight chain. These instabilities remain latent in
a straight infinitely long chain, because the bending of such a
chain would require an infinite energy. The role of the
charge-curvature interaction on the formation of the ground state
of \textit{closed} semiflexible  molecular chains was studied in
\cite{gaid-conf,gaid-conf-T,gaid-conf-L}.
 It was  shown that the coupling between charge carriers  and the bending degrees of
 freedom of the chain modifies the local bending rigidity of the semiflexible chain.  Due to the interaction between charge carriers and the
bending degrees of freedom the circular shape of the aggregate may
become unstable and the chain takes the shape of an ellipse or, in
general, of a polygon.
 There, the polygon structure is a result of the
self-consistent interaction between charge carriers and bending degrees
of freedom: extrema of the curvature and of the charge
density correlate: in the case of the softening charge-curvature
interaction  maxima of curvature and charge
density coincide, while in the case of the hardening interaction the minima of the curvature coincide with the maxima of the charge density. These results were obtained   by assuming that the  charge carrier
dynamics is coherent and the total charge is an integral of motion.

In this work we are interested in nonequilibrium shape transformations which may occur in closed filaments possessing two kinds of degrees of freedom: high frequency electromagnetically active degrees of freedom (in what follows we will call them excitons) and low frequency modes which are nonlinearly coupled with excitons. There are many examples of such systems. Probably, the best known model for excitations in a biological chain is the Davydov/Scott model for proteins. See Ref.~\cite{cruzeiro2016} for a recent review. In this model  the so-called Amide~I excitation consists of the stretching vibration of the $C=0$ bond of the peptide groups, which are linked by hydrogen bonds. The periodic forcing of the exciton modes can be provided by electromagnetic waves. The Amide I vibration in proteins has a frequency of 1665\,cm$^{-1}$ or about 50\,THz, i.e. between the near and mid-infrared spectrum.
There is abundant bibliography dealing with the interaction of infra-red radiation with proteins and in particular with the Amide I modes, including conformational changes and photoinitiated dynamics. See for example~\cite{Ganim2008,Barth2002} and references therein.  The interaction of the exciton with the bending degrees of freedom is also considered in a very similar model for Amide I excitations in crystalline acetanilide~\cite{cruzeiro2015}.
Lifetime of the excitations  is still a matter of research and debate. The lifetime of Amide I excitations in the acetanilide crystal has been experimentally determined in 2\,ps but it was also shown that the excitation energy is not dissipated until 35\,ps suggesting that   the Amide I excitation can be transformed into another more long-lived excitation~\cite{edler2002}. However, Amide I excitations in a protein~\cite{xie2002} survive much more, up to 500\,ps. Theoretical calculations in the Davydov/Scott model show that the exciton can travel along the protein in a few picoseconds time~\cite{cruzeiro2016}.

Another example  are DNA minicircles ~\cite{Bates2005,Bates13} and  circular plasmids adsorbed in a mica surface~\cite{Witz2008} with their far-infrared-active interbase hydrogen-bond breathing modes (characteristic frequency is $\sim 100\,\text{cm}^{-1}$ \cite{Chen95}~ and typical  life time is $\sim $10\,ps~\cite{Shih00})  which are nonlinearly coupled with torsional-acoustic modes~\cite{Golo05}.
An effective  control of the shape of the system by visible light is achieved by incorporating dye-monomers into liquid-elastic elastomers \cite{Woltman07,Ji12}. The  liquid-elastic elastomers are characterized by a strong coupling between the orientational order and mechanical strain. Under the action of linear polarized light nematic strips doped with azo-dyes controllably bend as monomers photoisomerize between their \textit{trans}- and \textit{cis}-states and reduce the degree of the orientational order in the elastomer~\cite{Camacho-Lopez04}.

The aim of the paper is to study shape transformations in driven and damped molecular chains. A \emph{generic} model of closed chain of weakly coupled molecular sub-units under the action of a spatially homogeneous time-periodic external electric field is studied. In contrast to the previous studies \cite{gaid-conf,gaid-conf-T,gaid-conf-L}, where the shape formation of the molecular chain was due to a charge-bending interaction, and therefore related to an \textit{equilibrium} phase transition, here we discuss the shape transformation of the molecular chain as a \textit{non-equilibrium} phase transition which occurs due to the energy pumping in the system.
The paper is organized as follows. In Sec. II we describe a model. Sec. III presents the stationary analytical solutions and discuss the stability issues. In Sec. IV we present the results of numerical simulations, demonstrating the excitation of different shape modes depending on the parameters. In Sec. V we
propose an analytical approach to the problem based on the Galerkin decomposition method, and compare the
analytical results  with the results obtained directly by
numerical simulations. Finally, Sec. VI presents some concluding remarks.

\section{The model}
We consider a simple phenomenological model of a polymer ring consisting of particles, labeled by an index $n$, and
located at the points $\vec{r}_n=\{x_n, \, y_n,\,z_n\}~(n=1\dots N)$. We
are interested in the case when  the array represents a closed
chain and so we impose the periodicity (closure) condition on the
coordinates $\vec{r}_n=\vec{r}_{n+N}$. Each unit $n$ is connected
with its two neighbors $n+1$ and $n-1$ by elastic bonds.
We will assume that the chain is inextensible: $|\vec{r}_n-\vec{r}_{n+1}|=a$ (the bond length $a$ is a constant which we in what follows put equal to $1$).
The change of the angle between  the bond  vectors
$\vec{t }_{n+1}=\left(\vec{r}_{n+1}-\vec{r}_n\right)~$
 and
 $\vec{t }_{n}=\left(\vec{r}_{n}-\vec{r}_{n-1}\right)~$is
 controlled by the bending potential which we take in the
 form
  \begin{eqnarray}\label{potbend}
U_b=\frac{K}{2}\,\sum_n\,\kappa^2_n\end{eqnarray}
 where $K$ is the elastic modulus of the bending
rigidity  (spring constant) of the chain,
 \begin{eqnarray}\label{kappan}
\kappa_n\equiv\,
\,|\vec{t }_{n+1}-\vec{t }_{n}|
\end{eqnarray}
 determines the curvature of the chain at
 the point $n$.
By using  the parametrization
\begin{equation}\label{param}\vec{t}_n=\Big(\cos\theta_n,\sin\theta_n \sin\phi_n,\sin\theta_n \cos\phi_n\Big),\end{equation}
 the local curvature can be presented in the form \begin{eqnarray}\label{kappan_angle}
 \kappa_n\approx
 \sqrt{(\theta_{n+1}-\theta_n)^2+\sin\theta_n\,(\phi_{n+1}-\phi_n)^2}.
 \end{eqnarray}
The angles $\theta_n$  and $\phi_n$ satisfy the relation
\begin{eqnarray}\label{periodic}
	\theta_{n+N}=2\pi+\theta_n,\qquad 	\phi_{n+N}=\phi_n
\end{eqnarray}
which stems from the periodicity (closure) condition.

The bending rigidity of the chain $K$  can be expressed as
\begin{eqnarray}\label{K}K = l_p\,k_B\,T,\end{eqnarray}
where $l_p$ is the persistence length in units of chain period,
$T$ is the temperature, and $k_B$ is the Boltzmann constant.

We consider the situation when the chain  adheres to  some surface. The interaction between the chain and the surface tends to orient all bond  vectors parallel to  the surface: easy-surface anisotropy interaction. It is assumed weak enough in order not to affect the internal and bending dynamics of the chain.   The surface of adherence is parallel to  a  $x-y$ plane and the easy-surface interaction has the form
\begin{eqnarray}\label{easy-surface}
U_w=\frac{w}{2}\,\sum_n\,(\hat{\vec{z}}\cdot \vec{t}_n)^2\equiv\nonumber\\
\frac{w}{2}\,\sum_n\,\sin^2\theta_n\,\cos^2\phi_n
\end{eqnarray}
where $\hat{\vec{z}}=(0,0,1)$  is a unit vector along the $z$-axis, and the parameter $w$ gives the intensity of the surface-chain interaction.

We assume that each particle represents a complex sub-unit of the chain which additionally to its position $\vec{r}_n(t)$, carries a high-frequency internal excitation which can be characterized by a complex amplitude, $\Psi_n(t)$. Examples of such internal modes are  Amide I vibrations in proteins or base-pair vibrations in DNA~\cite{peyrard} and optical excitations of dye-doped liquid-crystal elastomers~\cite{terentjev}, as commented in the introduction.
The Hamiltonian of the high frequency excitations (excitons) has the form
\begin{eqnarray}\label{ham-high}
H=\omega \,\sum_n |\Psi_n|^2+\frac{1}{2}\,J\,\sum_{n}\,|\Psi_{n+1}-\Psi_n|^2,
\end{eqnarray}
where $\omega$ is the energy of the excitation (the Planck constant is set equal to 1), the parameter $J$ characterizes resonance coupling between sub-units.
We will assume that the presence of the high frequency excitation at the site $n$ modifies locally the bending rigidity or, in other words, there is a coupling between excitons and bending degree of freedom of the form
\begin{eqnarray}
\label{ele-conf}
U_{ex-b}=\chi \,\sum_n\,|\Psi_n|^2\,\kappa^2_n,\end{eqnarray}
where the parameter $\chi$ characterizes the strength of the coupling (see Appendix A for details).
In what follows we restrict ourselves to the case of hardening exciton-curvature coupling: $\chi>0$.

Energy is pumped into the system by exciting it
with a time periodic external force
of amplitude $f$ and frequency $\Omega$. The corresponding
interaction energy is given by
\begin{eqnarray}\label{force}
U_f=-f\,\sum_n\,\Big(\Psi_n\,e^{i\Omega \,t}+c.c.\Big). \end{eqnarray}

 The equation of motion for the complex amplitude $\Psi_n$ has the form
\begin{eqnarray}
\label{dyneq}
i\dot{\Psi_n}&=&-i\alpha \,\Psi_n+\omega \,\Psi_n-J\,\left(\Psi_{n+1}+\Psi_{n-1}-2\,\Psi_n\right)+\nonumber\\
& &\chi \,\kappa_n^2\,\Psi_n+ f \,e^{-i\Omega t}
\end{eqnarray}
where $\alpha^{-1}$ gives the life time of the exciton.

For the sake of simplicity we will neglect inertia effects in the
dynamics of the mechanical subsystem and take the equations of
motion for the bending degrees of freedom in the form of
overdamped Lagrange-Rayleigh equations
\begin{eqnarray}
\label{Lagr-Rayl}
\frac{\partial {\cal F}}{\partial \dot{\xi }_n}=-\frac{\partial}{\partial \xi_n}
\Big(U_b+U_w+U_{ex-b}\Big),\nonumber\\ \xi_n\in (\theta_n,\phi_n)
\end{eqnarray}
where $\cal{F}$ is a dissipative function which is given by the expression
\begin{eqnarray}
\label{dissipat-f}
{\cal F}=\eta \,\frac{1}{2}\,\sum_n(\dot{\vec{r}}_{n+1}-\dot{\vec{r}}_n)^2\, ,
\end{eqnarray}
 the parameter  $\eta$  being is the relaxation time for the bending  degrees of freedom.
The dissipative function \eqref{dissipat-f} describes  internal
friction,
which is due to irreversible processes taking place within
the system. The function  \eqref{dissipat-f}  is the discrete version of the dissipation
function which is usually used in macroscopic elasticity
theory~\cite{landau1986}. In terms of the parametrization \eqref{param} the dissipation function \eqref{dissipat-f} takes the form
\begin{eqnarray}
\label{dissipat-f-angles}
{\cal F}=
\eta \,\frac{1}{2}\,\sum_n\Big(\dot{\theta }_n^2+\sin^2\theta_n\,\dot{\phi }_n^2\Big).
\end{eqnarray}

From Eqs. \eqref{Lagr-Rayl} and \eqref{dissipat-f-angles} we get
\begin{eqnarray}\label{equ-theta}
&\eta \,\dot{\theta_n}=-(K+2\chi \,|\Psi_{n-1}|^2)\,
(\theta_{n}-\theta_{n-1})+\nonumber
\\&
(K+2\chi \,|\Psi_n|^2)\,\Big(\theta_{n+1}-\theta_n+
\sin\theta_n\,\cos\theta_n\,(\phi_{n+1}-\phi_n)\Big)\nonumber\\
& -w\,\sin^2\theta_n\,\cos^2\phi_n,
\end{eqnarray}
\begin{eqnarray}\label{equ-phi}
\eta \,\sin^2\theta_n\,\dot{\phi_n}& =& -(K+2\chi \,|\Psi_{n-1}|^2)\,\sin^2\theta_{n-1}
(\phi_{n}-\phi_{n-1})\nonumber\\
& &+(K+2\chi \,|\Psi_n|^2)\,\sin^2\theta_n\,(\phi_{n+1}-\phi_n)\nonumber\\& &
+w\,\sin^2\theta_n\,\sin\phi_n\,\cos\phi_n
\end{eqnarray}

 To simplify notations it is convenient to use a rescaled complex amplitude,  transfer to a rotating frame of reference
 \begin{eqnarray}\label{rot}\Psi_n=\sqrt{\frac{K}{\chi }}\psi_n\,e^{-i\Omega t}\end{eqnarray}
and  measure all relevant variables in terms of the coupling strength $\chi$, defining $\bar{t}=\chi t$, $\bar{\omega }=\omega/\chi$, $
 \bar{\Omega }=\Omega/\chi$, $\bar{f}=f/\sqrt{\chi K}$, $\bar{J}=J/\chi$,$\bar{\alpha }=\alpha/\chi$, $\bar{\eta }=\eta \chi/K$, and $\bar{w}=w/K$.

 In the rescaled variables Eqs.~(\ref{dyneq}), \eqref{equ-theta}  and  \eqref{equ-phi} take the form
 \begin{eqnarray}\label{eq-psi-re}
i\dot{\psi_n}&=&-(\delta+i\alpha )\psi_n-J\left(\psi_{n+1}+\psi_{n-1}-
2\,\psi_n\right)+\nonumber\\
& &\kappa_n^2\psi_n+f,
\end{eqnarray}
\begin{eqnarray}\label{eq-theta-re}
&\eta \,\dot{\theta_n}=-(1+2\,|\psi_{n-1}|^2)\,
(\theta_{n}-\theta_{n-1})+\nonumber\\
&(1+2 \,|\psi_n|^2)\,\Big(\theta_{n+1}-\theta_n+
\sin\theta_n\,\cos\theta_n\,(\phi_{n+1}-\phi_n)\Big)\nonumber\\
& -w\,\sin^2\theta_n\,\cos^2\phi_n,
\end{eqnarray}
\begin{eqnarray}\label{eq-phi-re}
\eta \,\sin^2\theta_n\,\dot{\phi_n}& =& -(1+2 \,|\psi_{n-1}|^2)\,\sin^2\theta_{n-1}
(\phi_{n}-\phi_{n-1})+\nonumber\\
& &(1+2 \,|\psi_n|^2)\,\sin^2\theta_n\,(\phi_{n+1}-\phi_n)+\nonumber\\& &
w\,\sin^2\theta_n\,\sin\phi_n\,\cos\phi_n
\end{eqnarray}
where "bars" are omitted for simplicity and $\delta=\Omega-\omega$ is a detuning frequency.
 Our analytical approach is based on the assumption that the relaxation time $\eta$ is short ($\eta \rightarrow 0$) and therefore the bending degrees of freedom are slaved to the exciton ones.
In this case from Eqs. \eqref{eq-theta-re} and  \eqref{eq-phi-re} we get
\begin{eqnarray}\label{phi-theta-solu}\phi_n=\frac{\pi }{2},\nonumber\\
\theta_{n+1}-\theta_n=\frac{A(t)}{1+2\,|\psi_n|^2}
\;,\end{eqnarray}
where the function $A(t)$ is chosen in the form

\begin{eqnarray}
\label{A}
A(t)=2\,\pi \Big[\sum\limits_{n=1}^N\,\frac{1}{1+2\,|\psi_n|^2}\Big]^{-1}\end{eqnarray}
to satisfy the periodic boundary conditions given by Eq. (\ref{periodic}). By inserting Eq. (\ref{phi-theta-solu}) into Eq. (\ref{eq-psi-re}), we obtain that the exciton dynamics is governed by the following nonlinear integro-differential equation
\begin{eqnarray}\label{eq-q-int-dif}
& &i\dot{\psi }_n=-(\delta+i\,\alpha )\,\psi_n-\\
& & J\,\left(\psi_{n+1}+\psi_{n-1}-2\,\psi_n\right)+
\,\frac{A^2(t)}{\Big(1+2\,|\psi_n|^2\Big)^2}\,\psi_n+f\;. \nonumber
\end{eqnarray}

Note that according to Eq. \eqref{phi-theta-solu}, the azimutal angle is fixed, so motion is confined to a plane, as Fig. 2 shows.

\section{Solutions and stability}
Eq. (\ref{eq-q-int-dif}) has a spatially homogeneous solution
\begin{eqnarray}\label{hom}\psi_n(t)=\Psi,~~~\Psi=\frac{f}{\delta-\frac{1}{R^2}+i\,\alpha }\end{eqnarray}
when all subunits oscillate in phase and  the chain  has a circular shape with the curvature $\kappa_n= 2 \pi/N\equiv1/R$

To investigate the stability of the spatially homogeneous solution (\ref{hom}) (and the circular shape of the chain) we
assume that $\psi_n(t)=\Psi+\varphi_n(t)$, with
$
\sum_{n=1}^N\,\varphi_n(t)=0$,
and linearize Eqs. (\ref{A}) and (\ref{eq-q-int-dif})
with respect to $\varphi_n(t)$. As a result we obtain

\begin{eqnarray}
\label{eq-phi}
&i\dot{\varphi }_n=-(\delta+\frac{1}{R^2}+i \alpha )\,\varphi_n-J\,\left(\varphi_{n+1}+\varphi_{n-1}-2\,\varphi_n\right)+\nonumber\\
&\frac{2}{R^2}\,\frac{1}{(1+2|\Psi|^2)}\,\Big(
\varphi_n-2\,\,\Psi^2\,\varphi_n^*\Big)
\end{eqnarray}
\begin{figure}
\begin{center}
\includegraphics[clip,width=\columnwidth]{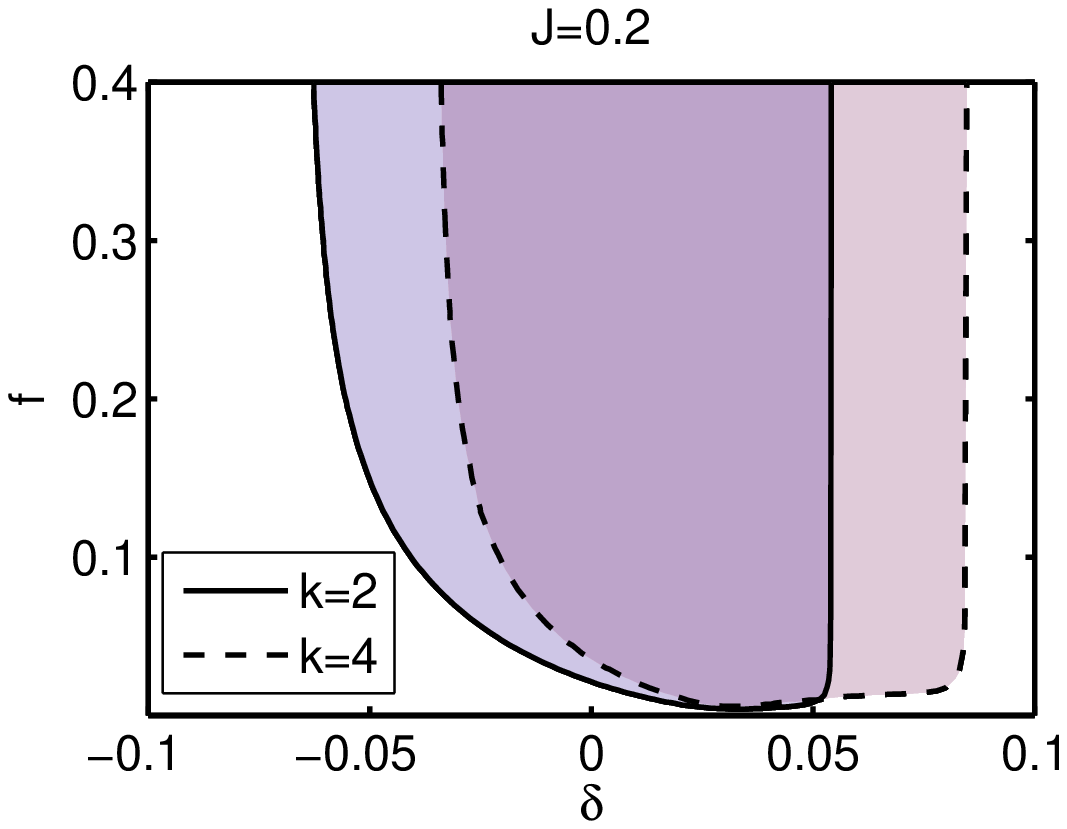} \\ \mbox{}\\
\includegraphics[width=\columnwidth]{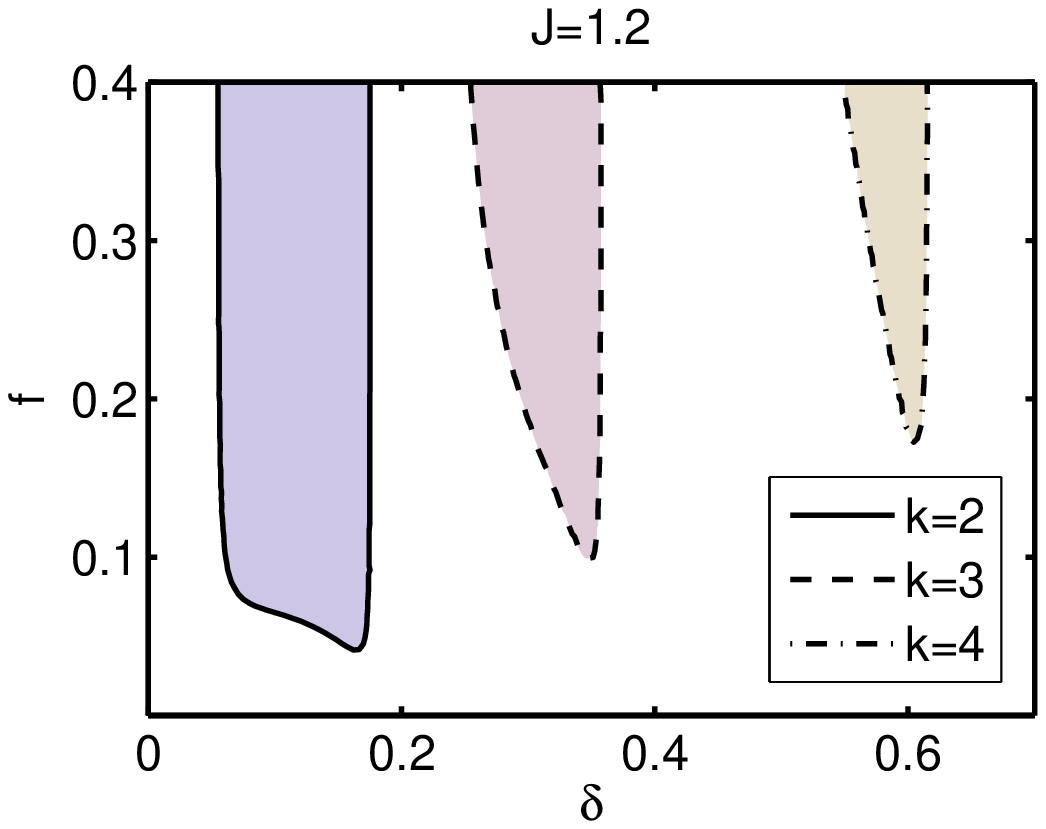}
\caption{Instability curves in $(\delta,f)$-space at which the $k$-th mode of the spatially homogeneous
solution destabilizes for $k$ = 2,3,4. The strength of resonance coupling $J$ varies in the
panels as indicated. We set weak losses, $\alpha=0.01$ and $\eta=0.01$. In the unshaded regions, spatially homogeneous states are stable.}
\label{fig_instab_anal}
\end{center}
\end{figure}

The stability is analyzed  by considering solutions of
the linear system (\ref{eq-phi}) of the form
\begin{eqnarray}\label{lin-phi}
\varphi_n(t)=\tilde{\varphi }\,\exp\{i\, \frac{2\pi \,k}{N}\,n-i\, z \,t\}\end{eqnarray}
where
$z$ is a complex frequency. Note that, in the linear approximation, the curvature of the chain is given by the expression
 \begin{eqnarray}\label{curv_linear}\kappa_n=\frac{1}{R}\,
 \Big(1-2\,\frac{\Psi \,\varphi_n^*+\Psi^*\,\varphi_n}{1+2\,|\Psi|^2}\Big)\;,\end{eqnarray}
 which means that in Eq.~(\ref{eq-phi})  the integer $k$ must satisfy  inequality $k\geq 2$  to fulfill the closure condition (\ref{closeness}); moreover, the  excitation of the $k$-th exciton mode corresponds to a $k$-gonal deformation of the chain: elliptical for $k=2$, triangular for $k=3$, etc.  Insertion of Eq.(\ref{lin-phi}) into Eq. (\ref{eq-phi}) leads to
\begin{equation}\label{zpm}
z_k=-i\alpha \pm\sqrt{\delta_k\Big(\delta_k+
\frac{8}{R^2}\,\frac{|\Psi|^2}{1+2\,|\Psi|^2}\Big)}\;,
\end{equation}
where $\delta_k=\Omega-\omega_k$ and
\begin{eqnarray}\label{omj}\omega_k=\omega+\frac{1}{R^2}+4\,J\,\sin^2\Big(\frac{\pi \,k}{N}\Big)\end{eqnarray}
is the frequency of the $k$-th exciton mode in the circular chain.
Direct inspection of Eq. (\ref{zpm}) shows that $Im(z_k)>0$ and the spatially homogeneous state (\ref{hom}) is unstable with respect to exciting the $k$-th exciton mode when $f\geq\,f_k$ and $\Omega \leq\omega_k$ ($\delta_k<0$), where
\begin{eqnarray}\label{instab_cond_j}
f_k^2=\Big(\alpha^2+\Big(\delta-\frac{1}{R^2}\Big)^2\,\Big)\,N_k\;,\end{eqnarray}
\begin{eqnarray}\label{Nk}
N_k=-\frac{1}{2}\,\frac{\alpha^2+\delta_k^2}{\alpha^2+\delta_k
(\delta_k+\frac{4}{R^2})}\;.\end{eqnarray}

We trace out instability curves of the spatially homogeneous state (\ref{hom})  in $(\delta,f)$-space.
 Equation (\ref{instab_cond_j}) defines the k-th mode stability curves in the $(\delta,f)$ plane. These results  demonstrate that for large  resonance coupling between subunits $J$, when
\begin{eqnarray}
\label{spit}
J>J_k=\frac{1}{N^2}\,\frac{4\pi \sqrt{1-\frac{\alpha^2 R^4}{4}} }{\sin\Big(\frac{\pi }{N}\Big)\,\sin\Big(\frac{\pi }{N}(2 k+1)\Big)}\,\end{eqnarray}
the  region of instability of the homogeneous solution splits into separate areas (see the bottom panel in Fig.(\ref{fig_instab_anal})) where the exciton modes with different $k$ grow indefinitely. However, for  $J< J_k$ the $k$- and $(k+1)$-th areas of instability  overlap and
 $k$-gonal  and $(k+1)$-gonal profiles  can be simultaneously stable for the
same parameter values. The top panel of Fig.~(\ref{fig_instab_anal}) presents the  situation when the homogeneous state is unstable simultaneously with respect to the modes $k=2$ and $k=3$.
\begin{figure}[t]
	\begin{center}
			\includegraphics[width=0.9\columnwidth]{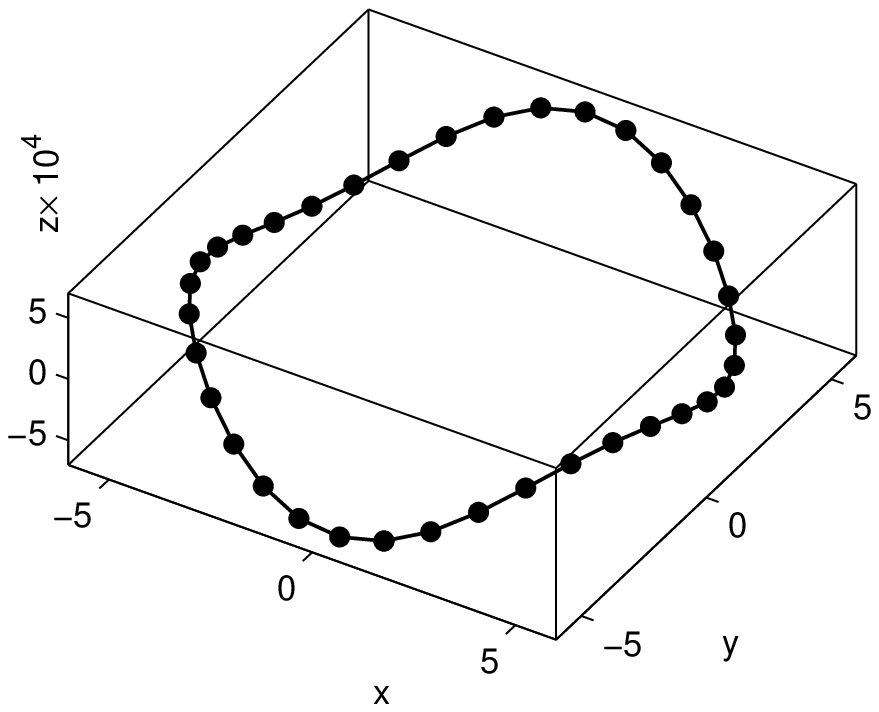}\\
			\includegraphics[width=0.9\columnwidth]{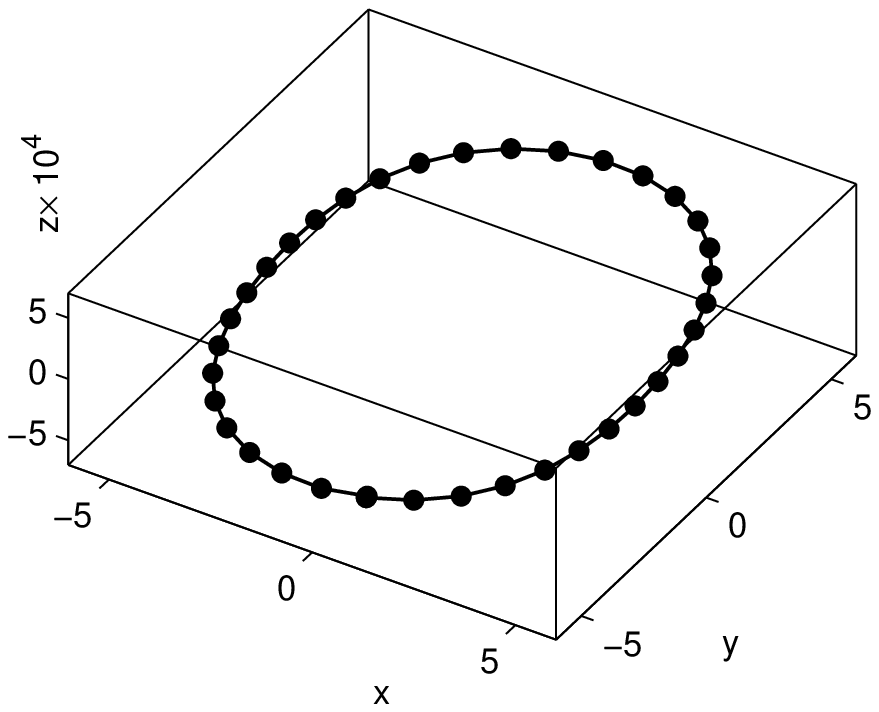}
		\caption{The top panel shows the initial 3D shape of the chain, and the bottom panel shows the stationary shape which is achieved by the chain in the presence of the same driving as in Fig.~\ref{fig_shapes}.}
\label{fig_shape_3D}
	\end{center}
\end{figure}

\begin{figure*}[t]
\begin{center}
\begin{tabular}{ccc}
\includegraphics[width=0.3\textwidth]{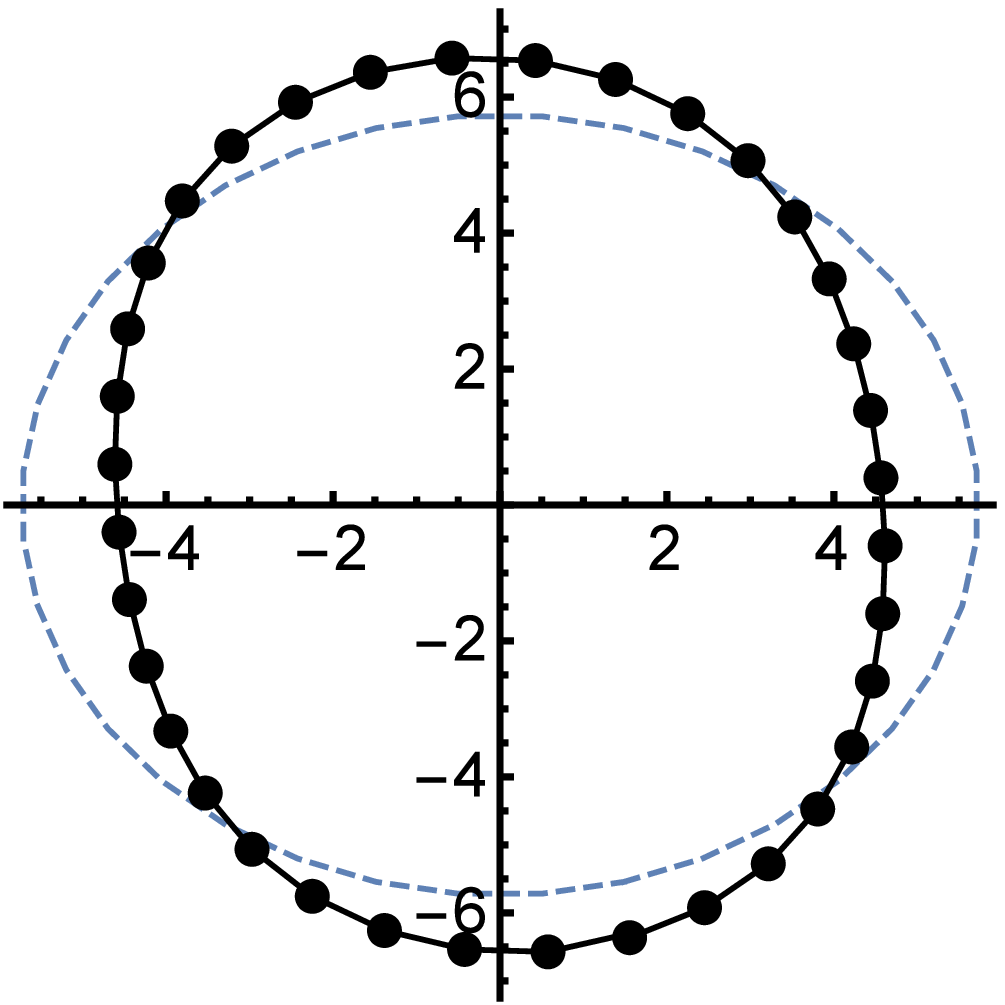} &
\includegraphics[width=0.3\textwidth]{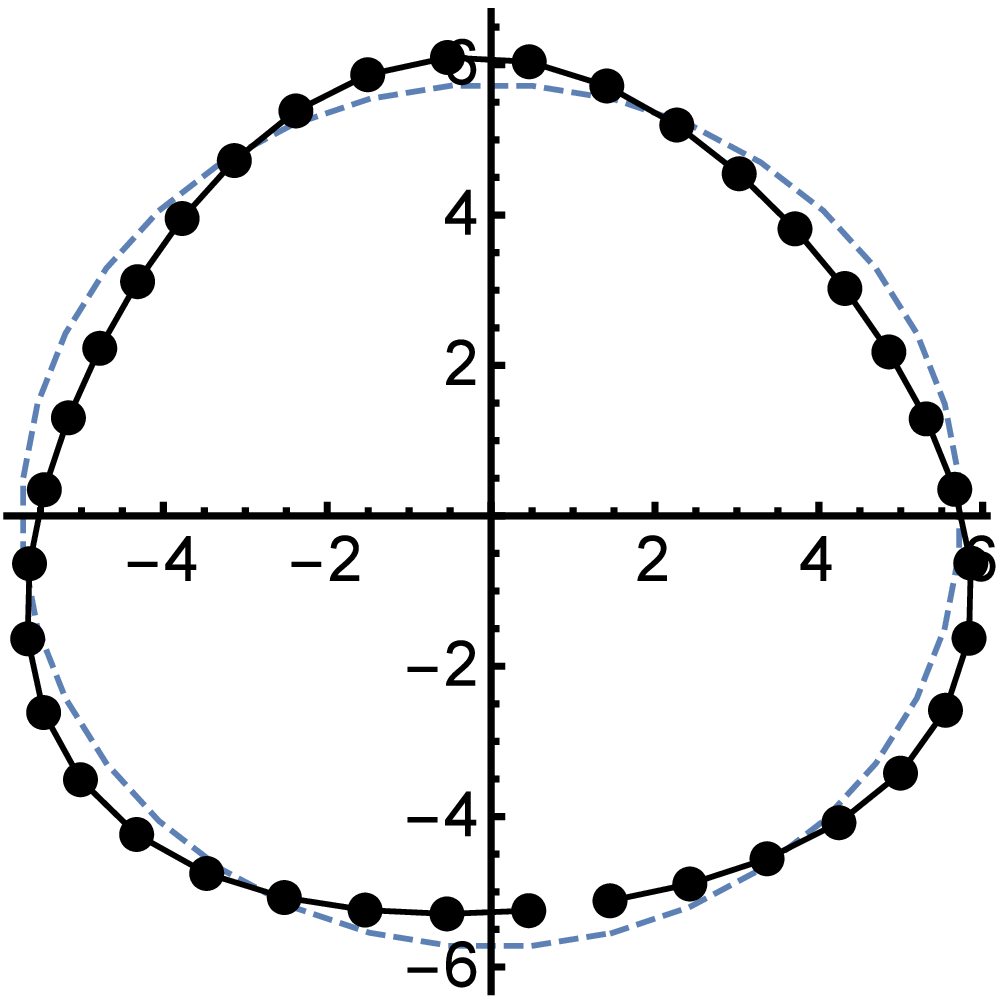} &
\includegraphics[width=0.3\textwidth]{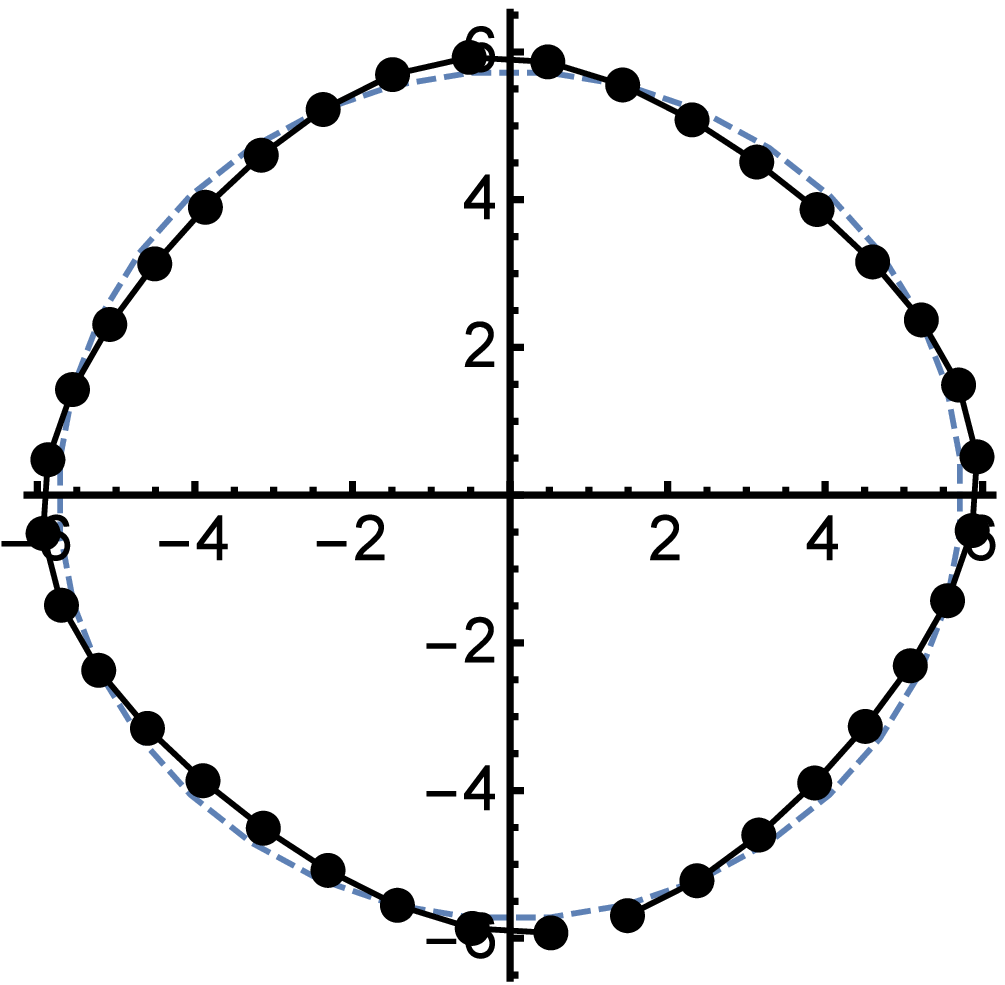} \\
\includegraphics[width=0.3\textwidth]{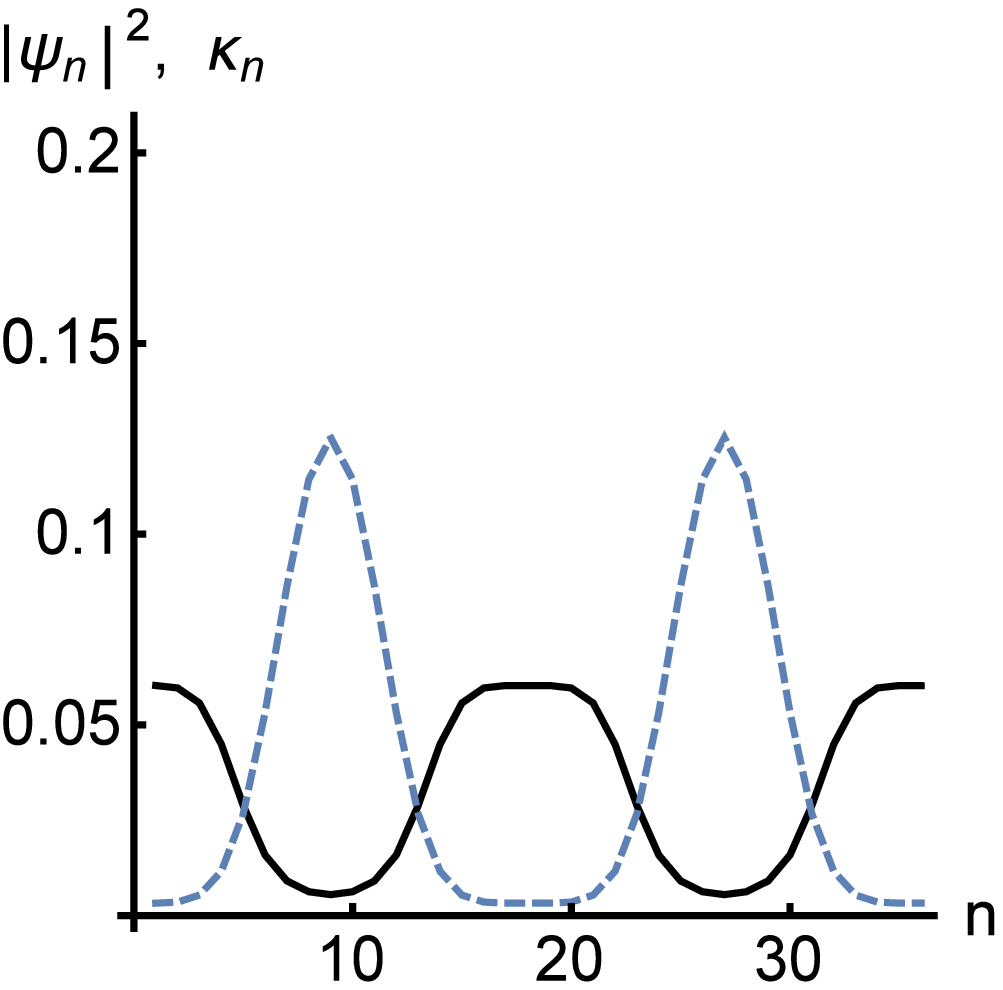} &
\includegraphics[width=0.3\textwidth]{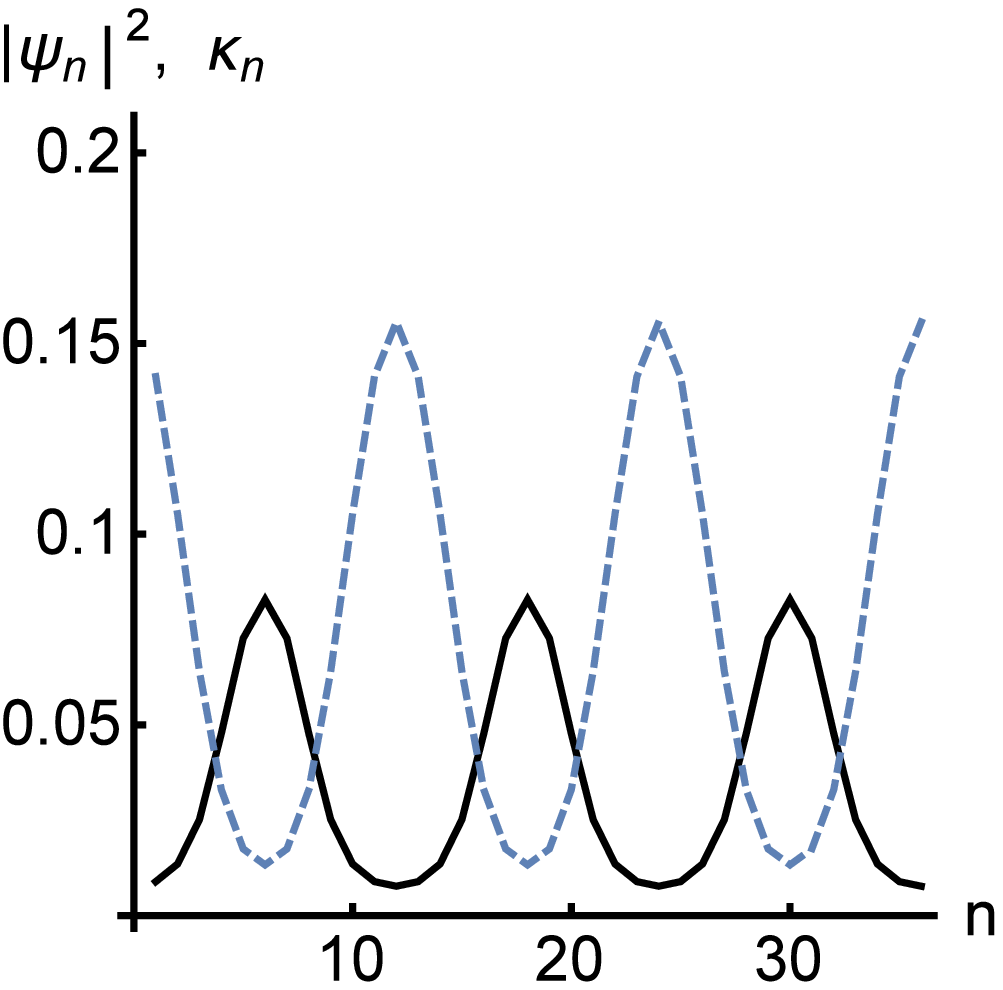} &
\includegraphics[width=0.3\textwidth]{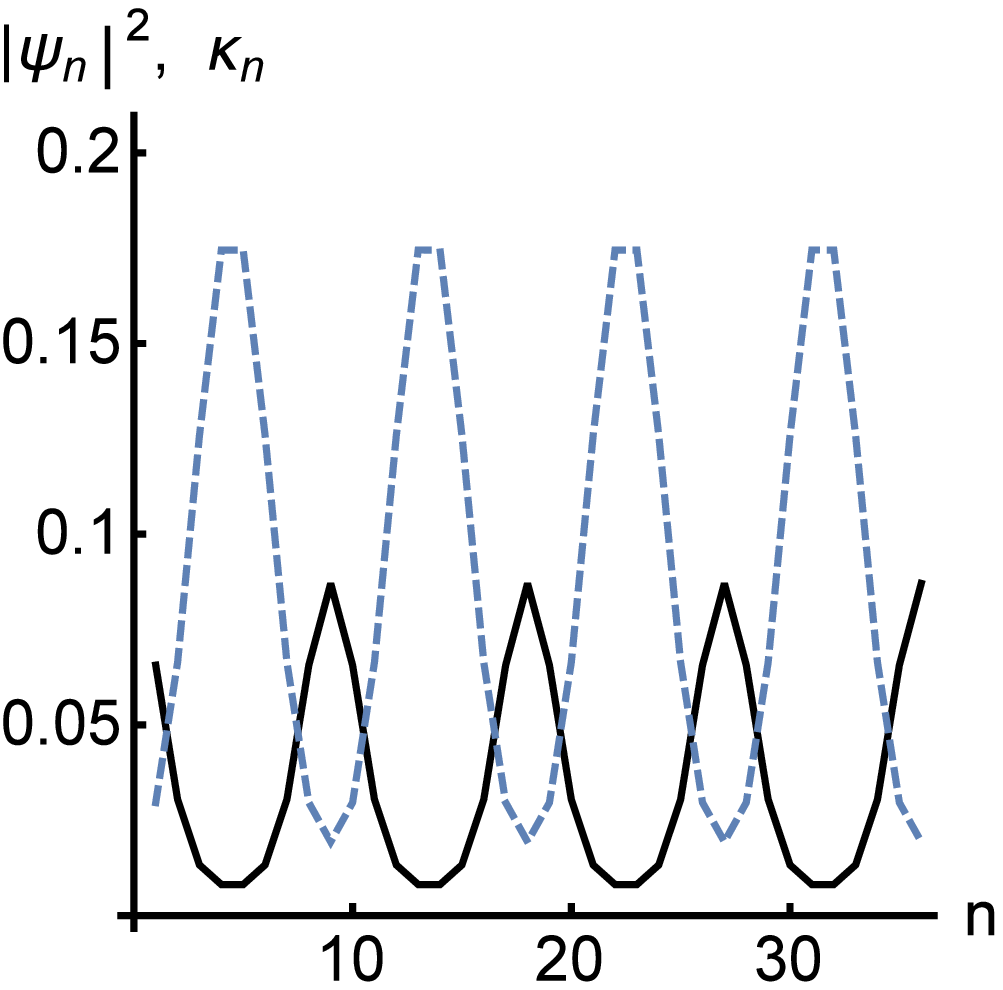}
\end{tabular}

\caption{\label{fig_shapes} Upper row: Mode shapes of index $k=2$ (left), $k=3$ (center) and $k=4$ (right), obtained in the area of parameters where the exciton modes are linearly unstable. Modes $k=2$ (elipselike) and $k=3$ (triangular) coexist and have been obtained for $f=0.01,J=0.2,\delta=0.03$. Mode $k=4$ (tetragonal) was obtained for $f=0.2,J=0.6,\delta=0.25$. Lower row: density of excitation energy distribution $|\psi_n|^2$ (dashed line, rescaled by a factor $10^{-1}$) and curvature variation $\kappa_n$  (solid line) along the chain.}
\end{center}
\end{figure*}

\section{Numerical results} To find the structures which appear as a result of such an instability we solved numerically Eqs.~(\ref{eq-psi-re}), \eqref{eq-theta-re} and (\ref{eq-phi-re}) for a chain of $N=36$ elements and different parameter sets. The starting configuration corresponded to the exciton subsystem in the ground state: $\psi_n(0)=0$
and initially, all the chain points were placed at (almost) symmetric points on the circle of an appropriate radius (the symmetry was broken by changing the local curvature of the chain by $0.1\%$).
The simulations were performed for
two initial seeds:
\begin{eqnarray}\label{seed}\theta_n=\frac{2\pi }{N}\,n+10^{-3}\,\sin\Big(\frac{2\,\pi \,k}{N}\,n\Big)\;,\nonumber\\
\phi_n=\frac{\pi }{2}-10^{-3} \cos\Big(\frac{2\,\pi \,k}{N}\,n\Big)\;,\qquad k=2,3\;\end{eqnarray}
Small losses are considered in all the cases, $\alpha=\eta=0.01$.
The coordinates of the subunits can be presented in the form
	\begin{eqnarray}\label{coord}x_n(t)&=&\sum_{m=1}^n\,\cos\theta_{m}(t)-
	\frac{1}{N}\,\sum_{m=1}^N\,(N-m-1)\,\cos\theta_m(t)\;,\nonumber\\
	y_n(t)&=&\sum_{m=1}^n\,\sin\theta_{m}(t)\sin\phi_m(t)-\nonumber\\&&
	\frac{1}{N}\,\sum_{m=1}^N\,(N-m-1)\,\sin\theta_m(t)\sin\phi_m(t)\;,
	\nonumber\\
	z_n(t)&=&\sum_{m=1}^n\,\sin\theta_{m}(t)\cos\phi_m(t)-\nonumber\\&&
	\frac{1}{N}\,\sum_{m=1}^N\,(N-m-1)\,\sin\theta_m(t)\cos\phi_m(t)\;,\end{eqnarray}
	where the last terms in the expressions for $x_n$, $y_n$  and $z_n$  fix   the center of mass of the chain at the coordinate origin. In terms of the parametrization (\ref{coord}) the closure condition reads
	\begin{eqnarray}\label{closeness}
	\sum_{m=1}^N\,\cos\theta_{m}(t)=\sum_{m=1}^N\,\sin\theta_{m}(t)\sin\phi_m(t)\nonumber\\
	=\sum_{m=1}^N\,\sin\theta_{m}(t)\,\cos\phi_m(t)=0.
	\end{eqnarray}

The results of the full scale 3D- simulations are shown on Figs. \ref{fig_shape_3D} and  \ref{fig_shapes}. It is seen that  while the initial state of the chain has a 3D-shape, the stationary  state has a flat shape parallel to the $x-y$-plane. We checked that the shape converges to a two-dimensional $x-y$ profile   even for a rather weak easy-surface anisotropy parameter: $w\sim 10^{-4}$.
The upper row of Fig.~(\ref{fig_shapes}) shows the elipselike ($k=2$), triangular ($k=3$) and tetragonal ($k=4$) distributions of the chain subunits obtained from numerical simulations. The lower row shows  the corresponding energy and curvature distribution along the chain. The elipselike and triangular modes have been obtained for the same forcing, coupling and detuning parameters, but different initial seeds, Eq. (\ref{seed}) with $k=2$ and $3$ respectively. This clearly indicates the existence of multistability predicted by the previous analysis, i.e. multiple $k$-gons, with different values of $k$, can be simultaneously stable for the
same parameter values. The energy and curvature distributions also demonstrate that the curve is more flat where the excitation density is maximal. Such a behavior is generic. The parameters used for obtaining the tetragonal shape were different from the ellipselike and triangular shapes (see caption).

We conclude that the polygon structure is a result of the
self-consistent interaction between excitations and bending degrees
of freedom: extrema of the curvature and of the excitation
density correlate: in the case under consideration when  the exciton-bending interaction locally hardens the chain stiffness
($\chi>0$)  the minima of the curvature coincide with the
maxima of the excitation density.

\section{Minimal model. Galerkin approach}
To gain some insight into the mechanism of shape transformations we will use a Galerkin approach by expanding the complex amplitude $\psi_n(t)$ into Fourier modes
\begin{eqnarray}\label{fourier}\psi_n=\sum_{k=-\frac{N}{2}+1}^{\frac{N}{2}}\,F_k(t) \,\exp\big({ i\,\frac{2\pi \,k\,n}{N}}\big)\;.\end{eqnarray}
Note that $F_1(t)=0$ in (\ref{fourier})  because the harmonics with $k=1$ can not contribute to the expansion (\ref{fourier}) due to the closure condition, Eq. (\ref{closeness}).

Lets us introduce an action functional $S=\int\, e^{2\alpha t}\,{\cal L}\;\mathrm{d} t$, where ${\cal L}$ is a Lagrangian function of the form

\begin{eqnarray}
\label{lagrange}
{\cal L}=\sum_{n}\Big\{\frac{i}{2}\Big(\dot{\psi }_n\,\psi_n^*-c.c.\Big)+\delta\,|\psi_n|^2-\nonumber\\
J\,|\psi_{n+1}-\psi_n|^2
-2\pi^2\,\Big[\sum_n\frac{1}{1+2\,|\psi_n|^2}\Big]^{-1}\}\;.\end{eqnarray}

Note that minimizing the action, $\delta S/\delta \psi_n^*=0$, leads to the evolution equation Eq.(\ref{eq-q-int-dif}).

By inserting Eq. (\ref{fourier}) into Eq. (\ref{lagrange}), expanding the Lagrange function (\ref{lagrange}) in terms of the amplitudes $F_k~(k\neq 0)$  up to the second order,  and carrying out summation over $n$, one can obtain an effective Lagrangian function in the form

\begin{eqnarray}
\label{lagrange-eff}
&{\cal L}=\frac{i}{2}\,\sum_{k}\Big(\dot{F}_k\,F_k^*-c.c\Big)+\Big(\delta-\frac{1}{R^2} \Big)\,|F_0|^2+ \nonumber \\
& \sum_{k\neq 0}\Big(\delta_k+\frac{1}{R^2} \Big)\,|F_k|^2
-\frac{2}{R^2}\,\frac{1}{1+2|F_0|^2}\,\sum_{k\neq 0}\Big(|F_k^2|-\nonumber\\
 & F_k\,F_{-k}\,\Psi^{* 2}-F_0^2\,F_k^{* }\,F_{-k}^*\Big)+f\,\Big(F_0+F_0^*\Big)\;.
\end{eqnarray}

From the action $S$ the evolution equations for the complex amplitudes $F_k$ can be obtained in the form
\begin{eqnarray}
\label{eqa0}
\lefteqn{i \dot{\Psi }=-\Big(\delta-\frac{1}{R^2} +i\,\alpha \Big)\,\Psi+}\nonumber\\ & &
\frac{4}{R^2}
\,\frac{1}{\Big(1+2|F_0|^2\Big)^2}\,\sum_{k\neq 0}\Big[\Big(|F_k|^2-F_k^*\,F_{-k}^*\,F_0^2\Big)\,F_0+\nonumber\\& &
+F_k\,F_{-k}\,\Big(1+|F_0|^2\Big)\,F_0^*\Big]-f\;,\end{eqnarray}

\begin{eqnarray}
\label{eqak}
\lefteqn{i\,\dot{F}_k=-\Big(\delta_k+\frac{1}{R^2}+i\,\alpha \Big)\,F_k+}\nonumber\\& &
\frac{2}{R^2}\,\frac{1}{1+2|F_0|^2}\,\sum_{k\neq 0}\Big(F_k-2\,F_0^2\,F_{-k}^*\Big)\;,~~k=2,3,\dots\end{eqnarray}
For each $k$ the set of equations (\ref{eqa0}),(\ref{eqak}) has two kinds of stationary solutions:\\
(1) $F_k=0$  for $k\neq 0$ and $F_0\equiv\Psi$ is given by Eq. (\ref{hom}).
In this state the excitation energy is {\em homogeneously} distributed along the {\em circular} chain. The stability of this state is discussed in the previous section (see Fig. (\ref{fig_instab_anal})).\\
(2) $F_k\neq 0$. In this state the chain is $k$-gonally deformed and the excitation energy is concentrated in the flat parts of the chain (as it is seen from Eq. (\ref{curv_linear}) a maximum of the curvature corresponds to a minimum of the excitation density and \textit{vice versa}). The stability regions in the $(\delta,f)$ parameter space of such kind of stationary state are presented  as shaded areas in Fig.  (\ref{fig_instab_anal}).

    The stationary state has a remarkable feature which distinguishes it from the first one. In contrast to the case of the circular chain when the amplitude of the spatially homogeneous component $|F_0|$ is a linear function of the pump intensity $f$ (see Eq. (\ref{hom})) in the stationary state of the second kind  this component does not depend on $f$
\begin{eqnarray}\label{Ns}|F_0|^2=-\frac{1}{2}\,\frac{\alpha^2+\delta_k^2}{\alpha^2+\delta_k
(\delta_k+\frac{4}{R^2})}\;\end{eqnarray}
and coincides with the threshold value $N_k$ at which   the spatially homogeneous excitation distribution becomes unstable with respect to excitation of the $k$-th exciton mode. At this point a Hopf bifurcation occurs and the amplitude of $k$-th mode for $f\rightarrow f_k$  evolves as
\begin{eqnarray}\label{Fk}F_k\sim\sqrt{f^2-f_k^2}\end{eqnarray}
with $f_k$ given by Eqs. (\ref{instab_cond_j}) and (\ref{Nk}).

\begin{figure}[t]
\begin{center}
\includegraphics[clip,width=\columnwidth]{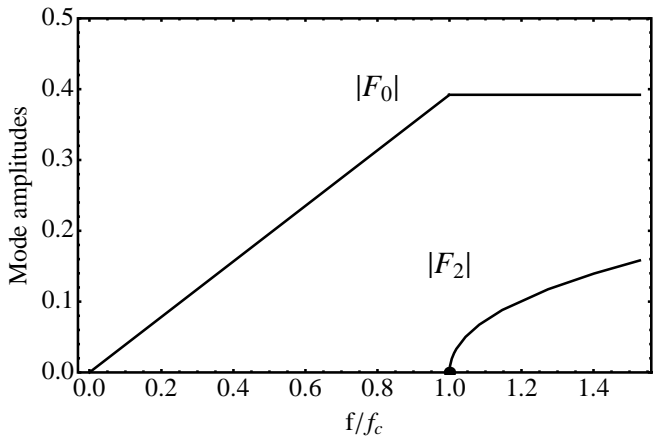}\\\mbox{}
\caption{Bifurcation diagram of the lowest harmonic modes $|F_0|$  and $|F_2|$  of the exciton wave function as a function of the normalized pumping strength $f/f_c$, where $f_c$ is the critical pump value at which the instability appears ($f_2$ in the case shown). The theoretical solutions (lines) were obtained in the framework of the minimal model,  while numerical results (symbols) correspond to full scale numerical simulations. Parameters are $N=36$, $J=0.2$, $\alpha=0.01$, $\eta=0.01$ and $\delta=0.03$.}
\label{fig_min_mod}
\end{center}
\end{figure}

This is illustrated in Fig. (\ref{fig_min_mod}) where the amplitude of the spatially homogeneous  component $|F_0|$ and the amplitude of the first nonvanishing Fourier mode $|F_2|$ are presented as a function of external force $f$. For $f\,<\, f_2$ the amplitude $F_0\sim F$ while $F_2=0$. At $f=f_2$ the Hopf bifurcation takes place and the behavior of the Fourier harmonics  qualitatively changes: $F_0$ remains constant and the amplitude of the second harmonics grows as Eq. (\ref{Fk}). To verify these results we carried full scale numerical simulations for the same set of parameters.
These results are shown in Fig.~(\ref{fig_min_mod}) with symbols. From this analysis, one can conclude that the minimal model gives a reasonable description of the dynamics of the system.

\section{Conclusions and discussion}

In this paper, we have studied the dynamics of closed chains of weakly coupled molecular subunits under the action of an spatially homogeneous time-periodic external field. Chains become flat because of  a weak interaction with a flat surface. We investigated the role of the exciton-curvature interaction on the formation of the shape and energy distribution of  closed semiflexible  molecular chains. The reported shape transformation of the molecular chain is the result of a non-equilibrium phase transition, which is mediated by an external driving (energy pumping) in the system. We have found that in the absence of driving the array takes a stable circular shape.
When the driving intensity exceeds some critical level the circular shape of the chain becomes unstable and the chain takes a polygonal shape. In the case under consideration, when the excitation locally hardens the bending rigidity of the chain, the excitation energy  is localized  in such places where the chain is more flat. The transition to the polygonal shape is due to a Hopf bifurcation.

\begin{acknowledgments}
Y.B.G. acknowledges partial financial support from a
special program of the National Academy of Sciences of
Ukraine, and is thankful to the Department of Applied
Mathematics and Computer Science and the Department of
Physics, Technical University of Denmark as well as the
University of Seville for hospitality. J.F.R.A acknowledges
Grant No. 2011/FQM-280 from CEICE, Junta de Andaluc\'{\i}a
Spain. J.F.R.A. and V.J.S.-M. acknowledge financial support
from Project No. FIS2015-65998-C2-2-P from MINECO,
Spain.

\end{acknowledgments}
\appendix*
\section{}
\label{appendA}
The aim of this appendix is to derive an explicit form  of the exciton-bending interaction for a generic 3D filament. To describe  the filament flexibility we use a discrete worm-like chain model. In the frame of this model  the  filament is considered as a chain of rigid links,length $a$, with vertices located at the points $\vec{r}_n=\Big(x_n,y_n,z_n\Big),~(n=1,...N)$. We are interested in the case when the chain is closed and so we impose  the periodicity condition on the coordinates:  $\vec{r}_{n+N}=\vec{r}_{n}$. We assume that each vertex represents a complex sub-unit of the polymer which additionally to its position $\vec{r}_n(t)$ carries an internal electromagnetically active degree of freedom which can be characterized  by  the complex amplitude $\Psi_n(t)$. We  consider a filament where electromagnetically active subunits are coupled to bending degrees of freedom. This coupling originates from the
change in the interaction energy (i.e. the van der Waals interaction,
isotropic part of multipole-multipole interactions, the
intermolecular exchange interaction, etc.) of the subunit with
all neighboring subunits in its transition to the excited
state, and can be written as follows:
\begin{eqnarray}\label{e-b}
E_{eb}=\sum_{n,n'}
V (|\vec{r}_n-\vec{r}_{n'}|)|\psi_n|^2,
 \end{eqnarray} where $V (|\vec{r}_n-\vec{r}_{n'}|)$
is the change of the interaction between
the $n$-th  subunit and the $n'$-th subunit when the former
occurs in the excited state.
In the next-nearest neighbor approximation the interaction  between the sub-units has the form
\begin{eqnarray}
\label{eb}
&E_{ex-b}=\sum\limits_{n=1}^{N}\Big\{V (|\vec{r}_n-\vec{r}_{n+1}|)+V (|\vec{r}_n-\vec{r}_{n-1}|)+\nonumber \\
 &V (|\vec{r}_n-\vec{r}_{n+2}|)+V (|\vec{r}_n-\vec{r}_{n-2}|)\Big\}\,|\Psi_n|^2=\nonumber\\
 &\sum\limits_{n=1}^{N}\Big\{2\,V(a)+V\Big(a\,\sqrt{4-\kappa^2_{n+1}}\Big)+\nonumber\\
 &V\Big(a\,\sqrt{4-\kappa^2_{n-1}}\Big)\Big\}\,|\Psi_n|^2 \,,
\end{eqnarray}
where the definition of the local curvature $\kappa_n$   \eqref{kappan} is used.

By assuming that $\kappa_n^2\ll 1$ and neglecting dispersion effects in the exciton-bending interaction,\textit{i.e.} $\frac{1}{2}\,(\kappa_{n+1}^2+\kappa_{n-1}^2)\approx \kappa_n^2$,  one can obtain approximately that
\begin{eqnarray}
\label{eb1}
E_{ex-b}=\sum_{n}\sum_{m=\pm 1,\pm 2}\,V(m\,a)\,|\Psi_n|^2+U_{ex-b},\nonumber\\
U_{ex-b}=\chi \,\sum_n\,\kappa_{n}^2\,|\Psi_n|^2\;.
\end{eqnarray}
Here the parameter
\begin{eqnarray}\label{chi}
\chi=-a\,\frac{d V(r)}{d r}\Big |_{r=2 a}
\end{eqnarray}
characterizes the intensity of the exciton-bending interaction.

%
\end{document}